\documentclass[conference,compsoc]{IEEEtran}
\usepackage{graphicx}
\usepackage{amssymb}
\usepackage{multirow}
\usepackage{xcolor}
\usepackage{adjustbox}
\usepackage{url}
\usepackage{comment}
\usepackage[para,online,flushleft]{threeparttable}

\ifCLASSOPTIONcompsoc
  \usepackage[nocompress]{cite}
\else
  \usepackage{cite}
\fi
\ifCLASSINFOpdf
\else
\fi
\begin{document}
%

\title{Chaos Engineering For Understanding Consensus  Algorithms Performance in Permissioned Blockchains}
\thispagestyle{plain}
\pagestyle{plain}

%

\author{\IEEEauthorblockN{Shiv Sondhi, Sherif Saad and Kevin Shi
}
\IEEEauthorblockA{University of Windsor\\
\{sondhis,shsaad, shi12z\}@uwindsor.ca}
\and
\IEEEauthorblockN{Mohammad Mamun}
\IEEEauthorblockA{National Research Council of Canada\\
mohammad.mamun@nrc-cnrc.gc.ca}
\and
\IEEEauthorblockN{Issa Traore}
\IEEEauthorblockA{University  of Victoria\\
itraore@uvic.ca
}}


%


\maketitle

\begin{abstract}
A critical component of any blockchain or distributed ledger technology (DLT) platform is the consensus algorithm. Blockchain consensus algorithms are the primary vehicle for the nodes within a blockchain network to reach an agreement. In recent years, many blockchain consensus algorithms have been proposed mainly for private and permissioned blockchain networks. However, the performance of these algorithms and their reliability in hostile environments or the presence of byzantine and other network failures are not well understood. In addition, the testing and validation of blockchain applications come with many technical challenges. In this paper, we apply chaos engineering and testing to understand the performance of consensus algorithms in the presence of different loads, byzantine failure and other communication failure scenarios. We apply chaos engineering to evaluate the performance of three different consensus algorithms (PBFT, Clique, Raft) and their respective blockchain platforms.  We measure the blockchain network's throughput, latency, and success rate while executing chaos and load tests.  We develop lightweight blockchain applications to execute our test in a semi-production environment. Our results show that using chaos engineering helps understand how different consensus algorithms perform in a hostile or unreliable environment and the limitations of blockchain platforms. Our work demonstrates the benefits of using chaos engineering in testing complex distributed systems such as blockchain networks.
\end{abstract}

%
\IEEEpeerreviewmaketitle

\section{Introduction}

 

\label{sec:intro}

\textcolor{black}{ A blockchain application is a distributed application running over a decentralized P2P network. The methods used to store and validate the blockchain network data enable data integrity, accountability, confidentiality, availability, and transparency. Blockchain introduced a new approach to creating an immutable distributed ledger for storing data. This ledger could also be tamper-resistant depending on the consensus algorithm used, as in the case of bitcoin and proof-of-work (PoW). }

\textcolor{black}{ The two terms, blockchain and distributed ledger technology (DLT), are often used interchangably. Blockchain is a type of DLT where transactions are recorded with an immutable cryptographic signature called a hash.  Blockchain networks store data in logical blocks that are chained together using these hashes. The data model is like a distributed linked list, where each block in the chain stores the address of the previous block, using a secure hashing approach. It is computationally expensive (almost impossible) to modify the block's data without being detected. The main difference between blockchain and DLT is the data model i.e. the structure used to capture the ledger's state. In a blockchain, the ledger is simply a linked list or a chain of blocks, but DLT could use any other data structure, such as a directed acyclic graph (DAG).  Any DLT is based on three components: a data model (e.g. linked-list, directed-acyclic graph), a transaction language, and a consensus protocol.}

\textcolor{black}{ 
The choice of consensus algorithm used in a given blockchain application significantly impacts the application's performance. In recent years, many consensus algorithms have been proposed for both public and private blockchain networks. Studying and comparing the performance of these consensus algorithms is very important. In this paper, we share our experience with analyzing the performance of consensus algorithms used in private blockchain networks.
}

\textcolor{black}{ 
In private or permissioned blockchain networks, the nodes and users are authorized before joining the network. These networks are suitable for applications in finance, logistics, healthcare, and other critical sectors. Several works in the literature have employed empirical analyses and modelling techniques to study the performance of consensus algorithms in blockchain networks. However, the performance of these algorithms, and their reliability in hostile environments, or in the presence of byzantine and other network failures is not well studied.  In our work, we apply chaos engineering and testing to study the performance of consensus algorithms in permissioned blockchain networks. The term chaos engineering was coined in 2014 by Bruce Wong at Netflix, while the practice of chaos engineering started in 2010 \cite{Basiri16, Tucker18, Torkura19}. In chaos engineering, we intentionally trigger faults into systems in production, to observe their behaviours in a faulty environment. This will help in implementing fault tolerance strategies that reduce downtime while increasing resiliency. The primary motivation for this approach is to overcome uncertainties prevalent in complex computer systems.
}

\textcolor{black}{We analyze the performance of three different blockchain platforms and consensus algorithms, by designing a lightweight blockchain application and implementing them separately on each platform. Then, using chaos testing principles, we perform load and chaos tests by introducing different user loads and network faults. We study the performance of the selected algorithms and platforms, under faulty conditions including network delay, packet loss, message corruption, crash failure, and Byzantine failure. To the best of our knowledge, our work is the first attempt to apply chaos engineering to study the performance and behaviours of consensus algorithms in blockchain networks.}

\textcolor{black}{The rest of this paper is structured as follows. Section \ref{sec:related} summarizes and discusses related works. In Section \ref{sec:consensus}, we introduce the consensus algorithms selected for analysis, and the metrics used to analyze them. Then, in Section \ref{sec:techniques}, we discuss our chaos testing approach, the blockchain application we designed for the test, and the blockchain platforms we used to deploy and test the application. In Section \ref{sec:performance}, we report the test results and discuss our observations and findings. Finally, Section \ref{sec:future} makes some concluding remarks and discusses our future work.
}

\section{Related Work}
\label{sec:related}
{\color{black}{Several works in the literature analyze the performance of blockchain consensus algorithms, using empirical analysis. Here, we briefly discuss some of the most recent work in the literature.}}

\begin{table*}[!htbp]
\caption{Comparing this work to the existing literature}\label{tab:lit}
\footnotesize
\centering
\begin{adjustbox}{width=0.75\textwidth}
\begin{threeparttable}

\begin{tabular}{|@{\vrule width0ptheight5pt\enspace}l|c|c|c|c|c|c|c|c|}\hline

\hfil\bf Paper& \bf Protocol Families &\bf Performance Metrics&\bf Load / Chaos Testing\\\hline

\hfil \cite{11_ampel19} &Paxos &TP\tnote{1}, L\tnote{2}, SR \tnote{3}, RU \tnote{4} &Load\\\hline
\hfil \cite{8_hao18} &PoW, BFT &TP, L &Load\\\hline
\hfil \cite{angelis18} &BFT, PoA &CAP &None\\\hline
\hfil \cite{ahmad21} &PoW, BFT, PoS, PoA, PoET &TP, L &Load\\\hline
\hfil \cite{14_papadis18} &Ethereum (i.e. PoW or PoA) &Block generation statistics &Load\\\hline
\hfil This work &BFT, PoA, Paxos &TP, L, SR &Both\\\hline

\end{tabular}
\begin{tablenotes}
\item[1]Throughput; \item[2] Latency;
\item[3] Success  Rate; \item[4] Node Resource Utilization
\end{tablenotes}
\end{threeparttable}

\end{adjustbox}
\end{table*}

In \cite{11_ampel19}, \textcolor{black}{Ampel et al.} used a performance benchmarking tool called Hyperledger Caliper \cite{21_caliper}, to measure the performance characteristics of a Hyperledger Sawtooth application. This application uses the Raft consensus protocol and the authors computed metrics like throughput, latency, success rate, and node resource utilization. {\color{black}{\textbf{\textit{Throughput}} is the number of transactions committed to the blockchain per unit time. \textbf{\textit{Latency}} is the amount of time it takes for a transaction to appear on the blockchain since when it was created. \textbf{\textit{Success Rate}} is the ratio of successfully committed blocks, to the total number of blocks created (including invalid blocks).}} These metrics are plotted against batch size (transactions per block) and the input workload. Noteworthy findings indicate that throughput increases linearly, and latency increases exponentially with batch size. Latency also increases exponentially with an increasing workload, while memory and CPU usage increase as well. 

In \cite{8_hao18}, \textcolor{black}{Hao et al. compared} the performance of Ethereum's \textcolor{black}{Proof-of-Work} (PoW) against Hyperledger Fabric's \textcolor{black}{Practical Byzantine Fault Tolerance} (PBFT). Average throughput and latency were once again used as comparison metrics. The results indicate that PBFT is better than PoW in terms of both metrics. {\color{black}{For smaller input workloads (around 100 transactions per second), PBFT was only slightly better than PoW, but as the workload increased, PBFT's performance grew far better than that of PoW.}} This is an indication of the poor scalability of the lottery-based PoW consensus. 
    
 \textcolor{black}{In \cite{angelis18}, Angelis et al. studied Aura and Clique - two variants of the Proof-of-Authority class of consensus algorithms; and classical PBFT, using the CAP (Consistency, Availability, Partition tolerance) theorem principles. The CAP theorem states that a distributed system  cannot achieve consistency and availability when the network is partitioned in a way that messages may be arbitrarily lost. In a blockchain network, consistency refers to all nodes having the same blockchain copy, and availability refers to the network's ability to accept new transactions. Through a qualitative analysis, the authors showed that Aura and Clique tend to prefer availability while PBFT prefers consistency.} 
 
\textcolor{black}{Ahmad et al. compared five different algorithms \cite{ahmad21} - PoW, PBFT, Proof-of-Stake (PoS), Clique and Proof-of-Elapsed Time (PoET) - on the basis of throughput and latency. The metrics were also measured when the number of network nodes was varied. They found that Clique and PoS experienced the minimum latency, followed by PoET, PoW, and PBFT. In terms of throughput they found that with up to 50 network nodes, Clique achieved the best throughput, followed by PoET and PoS. However, when the number of nodes increased beyond 50, Clique's throughput degraded. PBFT always had a very low throughput.}
    
\textcolor{black}{In \cite{14_papadis18} Papadis et al. used modelling techniques to analyze block generation statistics of a blockchain system. They compared the results using a blockchain application and a simulated model. They also analyzed the impact of stochastic components on the probability of attacks on the network. They used an Ethereum testbed for building the application. The authors found that the probability of a successful attack increases with increasing delay, and decreases with a higher number of transaction confirmations.}

{\color{black}{All the work discussed above used empirical analysis. The works in \cite{11_ampel19} and \cite{14_papadis18} measured the performance of one protocol. References \cite{8_hao18} and \cite{ahmad21} studied two and five protocols, respectively. While \cite{11_ampel19}, \cite{8_hao18} and \cite{ahmad21} used similar metrics, including throughput and latency, \cite{14_papadis18} analysed block generation statistics. \cite{11_ampel19} and \cite{8_hao18} conducted load tests although both used different methodologies. The former computed the metrics while varying the input transactions per second, while the latter varied the total number of transactions sent by a client to the server. In our experiments load is generated on the basis of the number of concurrent users interacting with the system. The experiments in \cite{ahmad21} and \cite{14_papadis18} vary the number of validator nodes on the network. The former goes over a larger range i.e. 50 to 250, whereas the latter tests only a 2-node and a 5-node scenario. In \cite{angelis18}, Angelis et al. conducted a qualitative analysis of three protocols using their algorithms and the CAP theorem.}

\textcolor{black}{Using the CAP theorem gives a different perspective on the characteristics of a protocol, however, this is not enough alone. The authors of \cite{angelis18} suggested that their analysis can be backed up by implementing the scenarios described in their paper, and computing various metrics including throughput, latency and scalability. Eventually, the CAP theorem can be used as a framework to analyze protocols, but metrics like throughput and latency are important to validate the model. The most common metrics used in previous works to measure
the performance of blockchain consensus algorithms consist of throughput and latency.  However, these two metrics alone, are not good enough performance indicators because these two metrics do not tell anything about the consistency of local chains nor do they say anything about the number of invalid, or rejected blocks.}

\textcolor{black}{Therefore, in our study we use throughput, latency, and success rate to compare the selected consensus algorithms. The success rate is taken as a ratio between the number of accepted blocks and the total number of blocks created (including the ones that were rejected). In addition, we consider two secondary metrics i.e. load tolerance and fault tolerance. Table \ref{tab:lit} highlights the difference between our proposed test strategy and the existing strategies in the literature. }

}

\section{Consensus Algorithms and Performance Metrics}
\label{sec:consensus}
There are close to a hundred consensus protocols used in blockchain and distributed ledger systems today \cite{encyclopedia18}. However, there is no single best protocol - the choice depends on network structure, topology, desired confirmation times, security and other factors. We focus on permissioned consensus algorithms and platforms suitable for healthcare, logistics, finance and other sectors that deal with sensitive and private users' information. The taxonomy from \cite{6_sadek20} was used to choose consensus algorithms which were considerably different from each other. Their structural and performance properties are most relevant here and are discussed below.

\begin{table*}[htbp]
\caption{\color{black}{Comparing the selected consensus protocols}}\label{tab:protocols}
\centering
\begin{adjustbox}{width=0.8\textwidth}
\begin{tabular}{|@{\vrule width0ptheight9pt\enspace}l|c|c|c|c|c|c|}\hline

\hfil\bf Protocol &\bf Family &\bf Platform & \bf Fault Tolerance &\bf Structure &\bf Underlying Mechanism\\\hline

\hfil PBFT &BFT-based &Hyperledger Sawtooth &BFT &Single Committee &Vote-based\\\hline
\hfil Clique &PoA-based &Ethereum's Rinkeby testnet &BFT &Single Committee &Leader-follower\\\hline
\hfil Raft &Paxos-based &Hyperledger Fabric &CFT &Single / Multiple Committee &Vote-based\\\hline

\end{tabular}
\end{adjustbox}
\end{table*}

\subsection{Consensus  Algorithms Selection}
Structural properties of consensus algorithms can be divided further into the following subcategories: 
\begin{enumerate}
    \item \textbf{Node type} - depending on the platform, a consensus algorithm may deal with multiple node types like full nodes (that store the entire blockchain locally), validator nodes, endorsers (which only validate transactions) and light clients (which verify new blocks without storing the entire blockchain locally).
    \item \textbf{Structure type} – Consensus protocols can use single or multiple committees to reach consensus i.e. a single group of validators generates each next block (as in PBFT, Tendermint and Clique), or multiple committees work independently. Both types can be static or dynamically changing. Furthermore, a single committee may be open or closed to new members, and can have implicit or explicit formation rules. Multiple committee mechanisms must have an overall topology (i.e. flat or hierarchical). {\color{black}{Raft normally follows a single committee structure, but when the network is partitioned, this splits into multiple flat committees. If any partition contains more than two-third of the participating nodes, it becomes the main committee and the others must follow its decisions (hierarchical topology).}} 
    \item \textbf{Underlying mechanism} – This refers to the core method of reaching consensus and can roughly be classified as either a \emph{lottery-based} (proof-of-work), \emph{vote-based} (BFT-based protocols) or \emph{coin-age-based} mechanism.
\end{enumerate}

The consensus protocols selected for this research - PBFT \cite{3_castro99}, Clique \cite{20_clique} and Raft \cite{5_ongaro14} - belong to the byzaninte fault-tolerant (BFT), proof-of-authority (PoA), and Paxos-based protocol families, respectively. BFT-based protocols are always byzantine fault-tolerant. This means that they reach a consensus even when a portion of the network's nodes send contradicting messages to different peers. Usually, BFT-based protocols follow multiple rounds of voting to achieve consensus - like PBFT - but this is not necessary. Many BFT-based protocols simply suggest improvements over PBFT, like reducing the number of voting rounds, etc. \textcolor{black}{PoA protocols are a popular class of non-incentivized protocols that store proof of each validator's identity to monitor and limit malicious activity. While PoA protocols are also byzantine fault-tolerant, they can reach better performance than BFT-based protocols due to lighter message exchanges. PoA protocols are best suited to scenarios where the validator set can be trusted, as is the case with Ethereum's Rinkeby, G\"{o}rli and Kovan testnets.} 

Finally, Paxos-based protocols provide improvements over the Paxos protocol proposed by Lamport in 1989. Raft (like Paxos) is not byzantine fault-tolerant but crash fault-tolerant. It is sometimes also referred to as a Proof-of-Capacity protocol \cite{encyclopedia18}. Table \ref{tab:protocols} summarizes the properties of the selected consensus protocols.

\subsection{Selected Performance Metrics}
Performance metrics are a way to quantify a system's performance. The performance properties of consensus protocols defined in \cite{6_sadek20} include throughput, latency, fault tolerance, and energy consumption. For our experiments, we selected three metrics (primary metrics) to directly measure aspects of the system; like throughput, latency and success rate. However, these metrics alone are not sufficient as they don't paint a holistic picture of system performance.   We employ chaos engineering techniques to study the effect of input traffic, failures, and combinations of failures on the blockchain system. Our focus is on input load and network failures. Therefore, we measure changes in the three primary metrics while changing the input workload, and adding network faults. The primary metrics we use to measure the performance of blockchain consensus algorithms and applications under different chaos conditions are:

\begin{itemize}
    \item \textbf{Write Throughput} - The number of transactions added to the blockchain per second. 
    
    \[TP = \frac{(total\ transactions\ added\ to\ chain)}{(total\ runtime)}\]

    \item \textbf{Average Write Latency} - The amount of time it takes for a transaction to appear on the blockchain, from when it was made. We are concerned with the average over all transactions. In the equation below,

    \[L = \frac{\sum_{tx=1}^{TX_{tot}} (T_{txCommitted} - T_{txCreated})}{TX_{tot}}\]
    
    Where \textcolor{black}{$TX_{tot}$ is the total number of transactions,
    $T_{txCommitted}$ is the timestamp when a given transaction is committed, and $T_{txCreated}$ is the timestamp when a given transaction is created (e.g. made by the user)}

    \item \textbf{Success Rate} - The ratio of the number of blocks successfully added to the blockchain to the total number of blocks created (includes invalid blocks). 
    
    \[SR = \frac{(total\ successfully\ added\ blocks)}{(total\ blocks\ created)}\]
\end{itemize}

\section{Chaos Engineering for Blockchain}
\label{sec:techniques}
\textcolor{black}{The majority of the work in the literature studied the performance of consensus algorithms under normal operational scenarios or in a fault-free environment. These studies are essential to understand how these algorithms behave and help to optimize and improve consensus protocols for blockchain networks. However, it is unrealistic to assume that blockchain applications and networks will continuously operate in a fault-free environment. It is critical to observe the performance of consensus algorithms and blockchain applications in faulty production environments. \textcolor{black}{This can be done using chaos engineering principles. Chaos engineering is a new system quality assurance practice that focuses on continuously testing complex distributed systems in production environments with stochastic faulty scenarios.}} 

\textcolor{black}{Any blockchain system can be divided into four abstract layers. These layers are the data model layer, consensus layer, execution layer, and application layer \cite{16_dinh17, 8_hao18}.  The data-model layer defines what data goes into a block - the data-structures and types. The consensus layer deals with finding consensus on the network and creating new blocks. The execution layer includes details of the runtime environment, which is used to execute smart contracts\footnote{For example, Ethereum's runtime environment is the Ethereum Virtual Machine (EVM).}. The application layer is the topmost layer and represents decentralized applications (Dapps) that use smart contracts and the blockchain to accomplish some business logic.}

\textcolor{black}{Consensus protocols are at the heart of the consensus layer. They are a well-defined instruction set that ensures all network nodes agree on the blockchain state (data-model layer). Therefore, the consensus and data-model layers are tightly knit. Changes in the network or data-model layers, like network delays, faulty nodes, corrupted messages, and block size, can affect the network's ability to reach consensus.}

\textcolor{black}{We designed several chaos testing scenarios to test the blockchain consensus algorithms and applications. We experimented with several users and transaction workloads to examine how the system behaves under different loads. In particular, we were interested in observing the point at which the system would crash, or its performance would severely degrade. We develop and execute various stochastic faulty scenarios to examine the system behaviours in a faulty production environment. The faulty stochastic scenarios included crash failure, Byzantine failure, and network communication failure. \textcolor{black}{A crash failure occurs when one or more blockchain validator nodes randomly crash.  Byzantine failure happens when one or more of the validator nodes randomly send contradicting messages over the network. Network communication failure occurs as a result of lost network packets or network delays. }}

\textcolor{black}{We used virtualization software and various blockchain platforms to build the blockchain networks for our testing. To conduct our chaos testing experiment, we used several chaos engineering tools. To deploy and run the nodes in the blockchain networks, we used Docker, a virtualization container\cite{Docker}. We used Locust \cite{26_locust}, an open-source distributed performance testing framework for writing and executing the load tests for different users and transactions. For executing the faulty stochastic scenarios, we used Pumba \cite{27_pumba}, a powerful chaos testing tool for injecting stochastic and random failures in Docker, such as crash failures and network failures. \textcolor{black}{We generated Byzantine failure by randomly corrupting outbound messages from a number of validator nodes.}}

\subsection{Blockchain Test Bed Construction}
The shortlisted consensus algorithms are available on different blockchain platforms. \textcolor{black}{For our experiments, we built an application on each platform, using the following business logic:} 
\begin{itemize}
    \item User A sends funds worth \emph{x} units to User B. 
    \item User A's account balance is decreased by \emph{x} units. 
    \item User B's account balance is increased by \emph{x} units.  
\end{itemize}

\textcolor{black}{This is a simple asset transfer application, however, depending on the use case, the platforms allow for much more functionality, including user registration and a fully functional web application. We selected a simple application in order to obtain results that were representative of the underlying protocols' performance - additional features would result in performance overhead.} Below is a brief overview of the selected platforms.

\subsubsection{\textbf{Hyperledger Sawtooth}}
Hyperledger Sawtooth \cite{22_sawtooth} is a modular framework, that separates the system's business logic from application-level procedures, making it easier for developers to work with. It supports dynamic consensus i.e. the ability to switch between consensus protocols in-between voting rounds, and pluggable consensus i.e. the ability to choose from a list of protocols. Sawtooth supports Go, Java, JavaScript and Python SDKs. 

The Sawtooth application was built using version 1.2.6 with the PBFT consensus model. Each node had four docker containers - a REST API endpoint, a consensus engine (PBFT), a validator, and a transaction processor. The default transaction processor was used in our experiments, which allows for the following types of transactions: creating an account with an initial balance, modifying and listing the value of an account, and listing the values of all accounts. \textcolor{black}{For each node, the REST API's port was exposed in the docker file and used for communication over the network.}

\subsubsection{\textbf{Go Ethereum}}
Go Ethereum (or geth) \cite{24_geth}, is an Ethereum client written in Go. Like other implementations of Ethereum, it resides on every node of the network and can run on the Ethereum mainnet as well as some testnets. As a result, it offers the Ethash protocol (Ethereum's PoW), IBFT \cite{ibft17}, and Clique. It works through a JSON-RPC API, and web3 libraries which allow developers to run, maintain, debug and monitor their nodes. Geth v1.10.3 was used in these experiments. 

The following steps were followed to build the Geth application: 
\begin{itemize}
    \item Create validator accounts (address, password, keys).
    \item Create the genesis block with Clique consensus, designated block creators, and account balances. 
    \item Compile each node's address into a static node list, which is shared amongst the validators. 
    \item Start all the nodes using the geth command. 
\end{itemize}

\subsubsection{\textbf{Hyperledger Fabric}}
Hyperledger Fabric \cite{25_fabric} is a permissioned DLT platform, with a modular and highly configurable architecture. The ledger is shared by organizations, each having its own peers and/or orderers. Fabric supports Javascript, Go and Python for its chaincode, and supports the Raft and Kafka consensus protocols. The transaction flow in a typical Fabric app is as follows:
\begin{itemize}
    \item The client sends a transaction to every organization, who validate it, and send back an endorsement if valid. 
    \item The client sends the transaction and endorsements to an orderer organization that runs the consensus protocol.
    \item Transactions endorsed by a majority are accepted. 
    \item Once ordered the transactions are sent to the organizations and committed by their peers. 
\end{itemize}

The application was built using Fabric 2.x and the following steps were followed to build it: 
\begin{itemize}
    \item Create certificate authorities and generate certificates for each organization using Docker and a Fabric binary. 
    \item Register orderers and peers with the organizations, and create crypto-material for them.
    \item Generate the genesis block and other channel artifacts. 
    \item Create the peers and orderers, along with their volumes and environments, using Docker. 
    \item Create the channel and join peers to it.
    \item Write the chaincode, install dependencies, package the chaincode, install it at the endorsing peers, and commit it if approved by a majority of the organizations. 
    \item Build the application using Node.js and Fabric API. 
\end{itemize}

Once the applications are built, their performance is compared using the metrics discussed in Section \ref{sec:consensus}. Further, each protocol is compared on the basis of load testing and chaos engineering, which is used to evaluate the fault tolerance of our applications. {\color{black}{Table \ref{tab:par-app}  shows the important parameters used in the tests, along with their values.

\begin{table}[h]
\caption{Application test parameters}\label{tab:par-app}
\centering
\begin{tabular}{|c|c|}\hline
\textbf{Parameter}&\textbf{Value}\\\hline
    Number of validators &6\\\hline
    Block size&10 tx/block\\\hline
    Baseline user load&250 or 50\\\hline 
    Load test user loads&250, 500, 1000, 1500\\\hline
    Locust workers&3\\\hline
    Users per second per worker&1, 2\\\hline
\end{tabular}
\end{table}
}}

Locust \cite{26_locust} is used to generate a constant, manageable load on the application, and the metrics are tracked over an entire test run. Locust interacts with the applications using HTTP requests and records the time for a response, the type of response (success or failure) and the total number of successful responses per second. For the load tests, the load is varied till the application crashes or performance degrades noticeably. For chaos testing, Pumba \cite{27_pumba} is used to generate network delay, loss, and message corruption for relevant network addresses. Pumba is used exclusively with Docker containers, therefore, for the Geth application (which does not use docker) each validator is created on a separate virtual machine and the traffic control (tc) tool within the Linux iproute2 package (used by Pumba under the hood) is used to introduce faults.

\section{Discussion Of Results}
\label{sec:performance}
{\color{black}{The baseline results presented in Table \ref{tab:baseline} show the throughput, latency, and success rate of each application calculated at a constant input load. Throughput and latency were also measured while varying the load and while adding faults to the blockchain network. The load test results are plotted in Figs.~\ref{fig:app-lt-pbft}-\ref{fig:app-lt-raft} and chaos test (fault tolerance) results in Figs.~\ref{fig:app-ft-pbft}-\ref{fig:app-ft-raft}. Table~\ref{tab:ft} presents the chaos test results by providing the average value for each metric (throughput and latency) while each network fault is being injected into the network.

\begin{table*}[tbp]
\caption{Blockchain applications: baseline performance results}\label{tab:baseline}
\centering
\begin{adjustbox}{width=0.8\textwidth}
\begin{tabular}{|@{\vrule width0ptheight9pt\enspace}l|c|c|c|c|c|c|c|}\hline
\hfil\bf Protocol&\bf Write Throughput (tx/s)&\bf Avg. Latency (ms)& \bf Success Rate&\bf User Count (Load)\\\hline
\hfil PBFT&50&1100&0.88&250\\\hline
\hfil Clique&27.3&49&1.0&250\\\hline
\hfil Raft&5.8&1850&0.98&50\\\hline
\end{tabular}
\end{adjustbox}
\end{table*}

Load is generated for the blockchain applications in terms of the number of users interacting with the app. In Table~\ref{tab:baseline}, a manageable load of 250 users was selected in order to get as stable results as possible. However, Raft could not deal with 250 users. This is down to how endorsement works in Hyperledger Fabric rather than due to the protocol itself. In order to endorse a transaction, peers first process the transaction and obtain the resultant ledger state, called the read set. After the transaction is accepted and ordered, before being committed, it is processed once again and the resultant state is called the write set. If the read and write sets do not match, the transaction is cancelled. This is not ideal for applications expecting large workloads because the state changes several times between generation of the read and write set. Companies like Boxer Construction Analysts and Robinson Credit Company have implemented independent solutions to deal with this issue \cite{18_fabric17}. Overall, PBFT seems to perform better in terms of throughput, and Clique in terms of average latency. Raft may perform better if Hyperledger Fabric is configured to deal with large loads.

\begin{figure}[tbp]
  \centering
  \includegraphics[width=0.75\hsize]{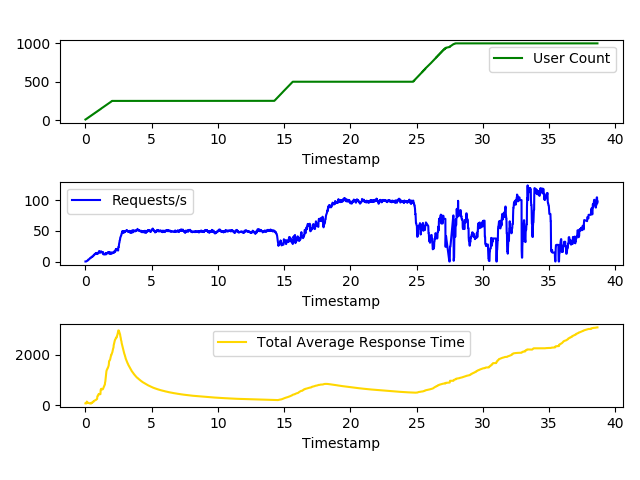}
  \caption{Load test: PBFT}
  \label{fig:app-lt-pbft}
\end{figure}

\begin{figure}[tbp]
  \centering
  \includegraphics[width=0.75\hsize]{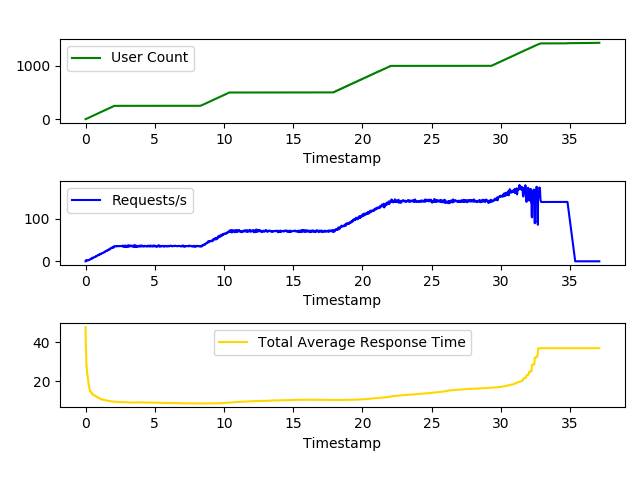}
  \caption{Load test: Clique}
  \label{fig:app-lt-cliq}
\end{figure}

\begin{figure}[tbp]
  \centering
  \includegraphics[width=0.75\hsize]{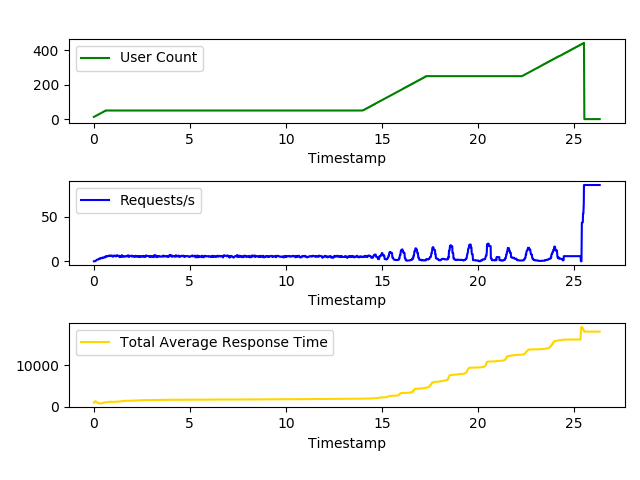}
  \caption{Load test: Raft}
  \label{fig:app-lt-raft}
\end{figure}

The load tests for each application were carried out until the application crashed, or performance degraded visibly. PBFT (Fig.~\ref{fig:app-lt-pbft}) did well till the load reached 1000 users, after which performance quickly degraded. PBFT's throughput and average latency fluctuate when the load is changing, but stabilize once the load stabilizes. The Clique application in (Fig.~\ref{fig:app-lt-cliq}) showed much better performance under load - it crashed once it reached 1500 users causing system performance to degrade. Raft (Fig.~\ref{fig:app-lt-raft}) performed the worst under load. As discussed, Hyperledger Fabric's inability to naturally handle large loads explains why performance is stable at lower loads but starts degrading/oscillating before even 250 users are spawned.

\begin{table*}[tbp]
\caption{Average performance metrics during chaos test}\label{tab:ft}
\centering
\begin{adjustbox}{width=1\textwidth}
\begin{tabular}{|@{\vrule width0ptheight9pt\enspace}l|c|c|c|c|c|c|c|c|}\hline

\hfil\bf Protocol&\bf Metric&\bf baseline&\bf delay (100ms)& \bf loss (15\%)&\bf delay+loss&\bf corrupted (50\%)&\bf corrupted+delay+loss&\bf paused (50\%)\\\hline

\multirow{2}{*}{PBFT}&Throughput(tx/s)&50&17.5&16.2&24.78&10.5&16.5&4.9\\\cline{2-9}
&Median Latency(ms)&18&4463&20.88&4475&2055&4513&Null\\\hline
\multirow{2}{*}{Clique}&Throughput(tx/s)&27.3&28&28.5&28.5&25.76&24&5\\\cline{2-9}
&Median Latency (ms)&6&105&6&110&7&103&Null\\\hline
\multirow{2}{*}{Raft}&Throughput (tx/s)&5.8&5&4.8&3.75&3.82&3.55&2.33\\\cline{2-9}
&Median Latency (ms)&1766&3150&3300&5100&6271&6430&18500\\\hline

\end{tabular}
\end{adjustbox}
\end{table*}

\begin{figure}[tbp]
  \centering
  \includegraphics[width=0.75\hsize]{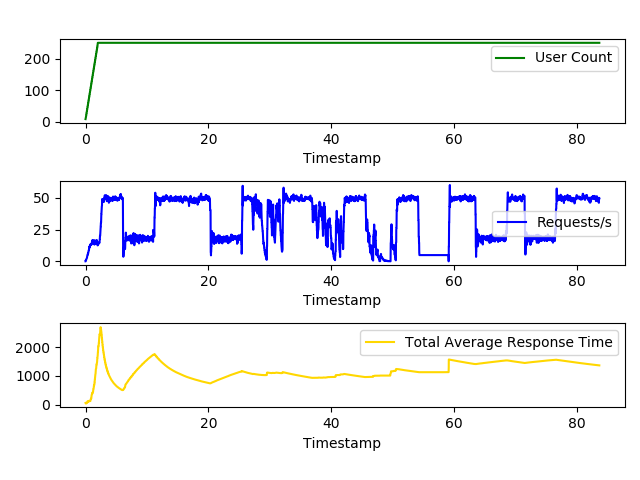}
  \caption{Fault tolerance: PBFT}
  \label{fig:app-ft-pbft}
\end{figure}

\begin{figure}[tbp]
  \centering
  \includegraphics[width=0.75\hsize]{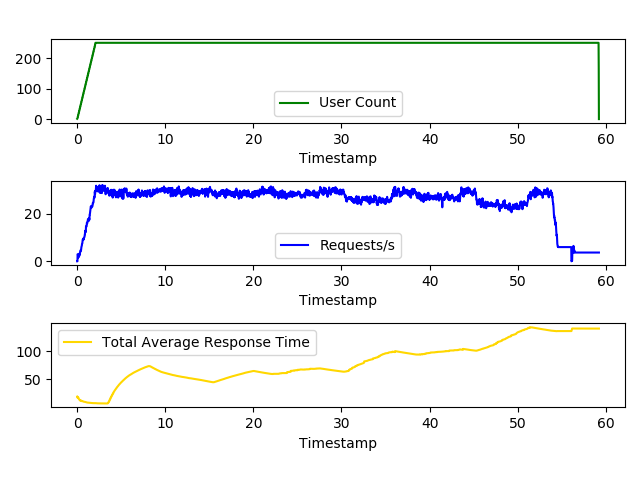}
  \caption{Fault tolerance: Clique}
  \label{fig:app-ft-cliq}
\end{figure}

\begin{figure}[tbp]
  \centering
  \includegraphics[width=0.75\hsize]{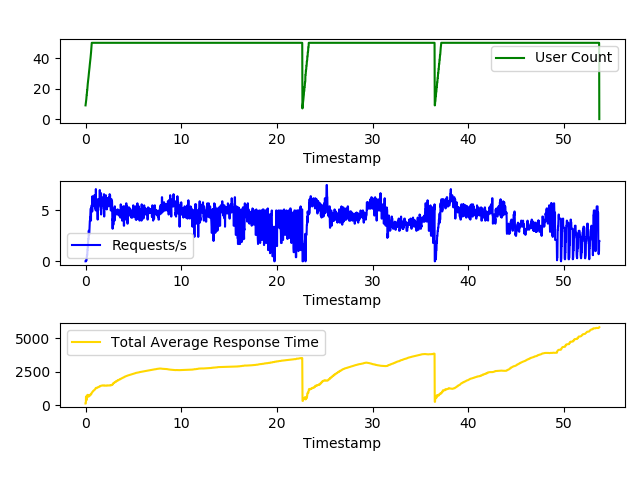}
  \caption{Fault tolerance: Raft}
  \label{fig:app-ft-raft}
\end{figure}

The chaos tests for each application were conducted at the same constant load as the baseline tests. The faults introduced during the test were (in order): delay, loss, delay and loss, corrupted messages from a single node, corrupted messages from half the network, corrupted messages (1 node) with delay and loss, corrupted messages (half network) with delay and loss, paused nodes. Here, corrupting outbound messages has a similar effect to byzantine activity as different nodes receive different messages. Similarly, pausing nodes is similar to crash failures. The metric values when certain network faults were injected are specified in Table~\ref{tab:ft}. Figs.~\ref{fig:app-ft-pbft}-\ref{fig:app-ft-raft} depict the entire test during which the faults were simulated consecutively. In these tests, after injecting each fault, the network was returned to normal conditions for an equal period of time, before injecting the next fault. This can be observed in Fig.~\ref{fig:app-ft-pbft} where throughput returns to the baseline periodically. The throughput in these plots can be compared to the throughput in Table~\ref{tab:ft}. However, the latency in Table~\ref{tab:ft} refers to the median latency at each instant during the test, while the latency in Figs.~\ref{fig:app-lt-pbft}-\ref{fig:app-ft-raft} represents a running average of the latency throughout the entire test run. 

Entries with 'Null' in Table~\ref{tab:ft} signify that no data is available for that period of the test. This is usually accompanied by a few short spikes where the latency metric degrades heavily. While the median response time (median latency) may remain relatively low during each spike, the maximum response time shoots up. For instance, when half the network was paused, the maximum response time degraded to 300000 ms in PBFT and 28000 ms in Clique. Apart from these short spikes, there is no data for latency during the periods in question. Pausing half the network nodes has the most dramatic effect on performance compared to other faults. One noteworthy observation is that network faults affect PBFT's throughput drastically, but have very little effect on Clique's throughput. On the other hand, Clique's and PBFT's average latency does not change drastically, whereas Raft's average latency is continuously degrading as network faults are added and removed from the network. 

The Fabric application could not handle the test very well and crashed thrice, hence the drops in the plots of Fig.~\ref{fig:app-ft-raft}. In fact, this figure consists of three separate tests whose results were combined together. The throughput plot for Raft looks like it fluctuates a lot, but this is due to the scale of the y-axis and in reality, the extremes are not separated by much at the baseline. Similar to Clique, the faults affect Raft's latency more than its throughput. It can also be seen that Raft handles network delay or loss well, compared to other faults.}}

\section{Conclusion}
\label{sec:future}
In this paper, we summarized our experiences in applying chaos engineering principles to blockchain consensus algorithms and applications. Using chaos engineering, we can observe the performance of consensus algorithms and blockchain applications in faulty production environments. In particular, to find the correlation between stochastic network faults and system performance.  We observed how the performance of the selected consensus algorithms changes as a result of user/transaction load and stochastic failures. It is clear that the choice of consensus algorithm affects system performance. 

We found that Clique was able to handle load better than PBFT and Raft, and also maintained its throughput in faulty environments. However, PBFT showed a better throughput overall while Raft performed the worst. In addition, our experiments with blockchain platforms show that the choice of blockchain platform plays an important role too. This indicates that if two different blockchain platforms use the same blockchain algorithm or protocol, we should not assume that they will have similar performance.  This can be due to restrictive architecture as in some platforms or extra effort as in Hyperledger Fabric (Raft). Eventually, it is the choice of consensus protocol as well as the platform that decides the performance of a blockchain system. 

In the future, we plan to extend our chaos testing scenarios to design a complete chaos test suite for blockchain applications. We will investigate the reliability of more complex blockchain applications (functional scalability) in the presence of failures. Investigate the impact of failure on geographic scalability. Finally, we are interested in investigating the overhead introduced by different blockchain platforms, particularly the platforms that use the same consensus algorithms or protocols.

\ifCLASSOPTIONcompsoc
  \section*{Acknowledgments}
\else
  \section*{Acknowledgment}
\fi

The authors would like to thank Canada NRC and the Artificial Intelligence for Logistics Program.
This project was supported in part by collaborative research funding from the National Research Council of Canada’s Artificial Intelligence for Logistics Program. This research was partially supported by the Scotiabank Global Trade Transactions Initiative administered by the University of Windsor’s Cross-Border Institute and Mitacs. We thank our colleagues from Scotiabank and Cross-Border Institute, who provided insight and expertise that greatly assisted the research. However, they may not agree with all of the interpretations/conclusions of this paper.



%


\end{document}